\documentclass[12pt]{emulateapj}

\slugcomment{}

\shorttitle{UB photometry of old open clusters.}
\shortauthors{Carraro et al.}

\begin{document}

\title{UB CCD photometry of the old, metal rich, open clusters NGC~6791, NGC~6819 and NGC~7142}

\author{G. Carraro\altaffilmark{1}}
\affil{European Southern Observatory, Alonso de Cordova 3107, 
Casilla 19001, Santiago 19, (Chile)}
\email{gcarraro@eso.org}

\author{A. Buzzoni}
\affil{INAF - Osservatorio Astronomico di Bologna, Via Ranzani 1, 40127 Bologna (Italy)}
\email{alberto.buzzoni@oabo.inaf.it}

\author{E. Bertone}
\affil{INAOE - Instituto Nacional de Astrof\'\i sica Optica y Electr\'onica, 
	      Calle L.E. Erro 1, 72840 Tonantzintla, Puebla (Mexico)}
\email{ebertone@inaoep.mx}

\and

\author{L. Buson}
\affil{INAF - Osservatorio Astronomico di Padova, Vicolo Osservatorio 5,
             35122 Padova (Italy)}
\email{lucio.buson@oapd.inaf.it}

\altaffiltext{1}{On 
leave from Dipartimento di Fisica e Astronomia, 
Universit\`a di Padova, Italy}

\begin{abstract}
We report on a UV-oriented imaging survey in the fields of the old, metal-rich 
open clusters, NGC~6791, NGC~6819 and NGC~7142. With their super-solar metallicity 
and ages $\gtrsim 3$-8 Gyr, these three clusters represent both very near and ideal 
stellar aggregates to match the distinctive properties of the evolved stellar 
populations, as in elliptical galaxies and bulges of spirals. Following a first
discussion of NGC~6791 observations in an accompanying paper, here, we complete
our analysis, also presenting for NGC~6819 and NGC~7142 the first-ever $U$ CCD 
photometry.
The color magnitude diagram of the three clusters is analyzed in detail, 
with special emphasis to the hot stellar component. We report, in this regard,
one new extreme horizontal-branch star candidate in NGC~6791.
For NGC~6819 and 7142, the stellar luminosity function clearly points to
a looser radial distribution of faint lower Main Sequence stars, either as a consequence 
of cluster dynamical interaction with the Galaxy or as an effect of an increasing fraction 
of binary stars toward the cluster core, as actually observed in NGC~6791 too.
Compared to a reference theoretical model for the Galaxy disk, the analysis of the 
stellar field along the line of sight of each cluster indicates that a more centrally 
concentrated thick disk, on a scale length shorter than $\sim 2.8$~kpc, might better 
reconcile the lower observed fraction of bright field stars and their white-dwarf 
progeny.
\end{abstract}



\section{Introduction}
Old open clusters are widely recognized as valuable tools to study the stellar
population of the Galactic thin disk \citep{bragaglia06,carraro07} and, at the same time,
as important benchmarks to probe stellar structure and evolution theories.
Recently, much attention has been paid to the evolution of stars along the red giant
branch (RGB), and the role of metallicity as main driver of mass loss \citep[e.g.][]{vanloon06,origlia07} 
and possible origin of extended blue horizontal branch (BHB) stars. In this context old, metal rich,
open clusters are ideal targets and, among these, NGC~6791 certainly stands out 
for its conspicuous population of blue horizontal-branch (BHB) stars \citep{kaluzny95,brown06}, 
and a wealth of white dwarfs (WD) \citep{bedin08}. However, the lack of high-quality UV 
photometry, particularly in the $U$ band, prevented so far a full characterization of the 
BHB component both in terms of completeness and UV properties.

This is the main scope of the present study, in which we present accurate wide-field
$UB$ photometry across the cluster NGC~6791. This photometric material provided the reference 
for \citet{buzzoni12} to characterize the UV properties of this cluster leading to conclude 
that it can robustly be considered as a nearby proxy of the elliptical galaxies
displaying a strong UV-upturn phenomenon. However, a detailed description of the photometric
data, their reduction and calibration, was deferred to the present paper.
Together with NGC~6791, we are going to present here $UB$ photometry for two additional
old, likely metal-rich, open clusters, namely NGC~6819 and NGC~7142, for which CCD $U$ photometry
is not available so far.
The main aim is to describe the color-magnitude diagram (CMD) in these pass-bands and, in case,
assess the possible presence of BHB candidate stars.

\subsection{NGC~6791}

Besides NGC~188, this object is the only relatively close system known to contain a
sizable fraction of sdB stars \citep{landsman98}. Located less than 5~kpc away 
\citep{carraro99,carraro06,carney05}, it stands out as a treasured
``Rosetta Stone'' to assess the UV emission of more distant ellipticals \citep{buzzoni12}. 
Though the first detailed study of NGC~6791 goes back to the work of \citet{kinman65},
its truly peculiar hot-HB content has indeed been recognized a few decades later, 
when \citet{kaluzny92} (hereafter, KU92) as well as \citet{kaluzny95}  (hereafter, KR95) verified that 
it hosts a significant fraction of sdB/O stars. Later, \citet{yong00} interpreted 
these hot sources as extreme horizontal-branch (EHB) stars with $T_{\rm eff}$ in the range 24--32\,000~K, as also 
confirmed by ground and space-borne (UIT and HST) observations \citep{liebert94,landsman98}. 

Its old age, about 8~Gyr, has recently been confirmed by \citet{anthony07} 
using $vbyCaH\beta$ CCD photometry, while a recent estimate of metallicity 
(i.e.\ $[Fe/H]\sim +0.40$) has been provided by \citet{carraro06}, \citet{origlia06}, 
and \citet{gratton06}, relying on high-resolution spectroscopy.

\subsection{NGC~6819}

A first hint of a relatively old age for this cluster dates almost 30~years ago,
from the photographic studies of \citet{lindoff72} and \citet{auner74}, which
compared the turnoff and red giant branch location relative to the CMD of 
the evolved system M~67. More recent and accurate age estimates from deep $BVI$ CCD 
photometry \citep{carraro94,kalirai01,rosvick98,warren09} better agree around a value 
of $\sim 3$~Gyr. No $U$ photometry has been published so far for this cluster.

Chemical abundances from high-resolution spectroscopy of red-clump stars in the cluster
have recently been presented by \citet{bragaglia01} and \citet{warren09}, suggesting 
a value of $[Fe/H]~= +0.09$. This consistently agrees with the original estimate by 
\citet{twarog97}, based on Str\"omgren photometry.

\subsection{NGC~7142}

The similarity of the NGC~7142 CMD with that of the old open clusters NGC~188 
and M~67 has been pointed out by \citet{vdb62}. Specific $BV$ CCD photometry has been 
carried out by \citet{crinklaw91} pointing to an age of 4-5~Gyr for this cluster, 
actually intermediate between that of M67 and NGC~188. This estimate matches both the 
very early observations of \citet{vdb62} and the more recent results of \citet{carraro94}. 
As far as metallicity is concerned, \citet{jacobson07,jacobson08} ascribe 
to NGC~7142 a moderately super-solar metal content, with $[Fe/H] = +0.14$.
The only modern CCD study of this cluster is from \citet{janes11}, in the $BVI$
pass-bands, and supports previous estimates for age, distance and reddening.

\begin{figure}
\includegraphics[width=0.96\hsize]{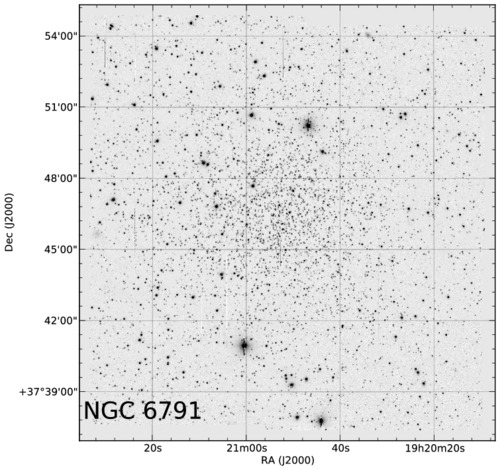}
\includegraphics[width=0.96\hsize]{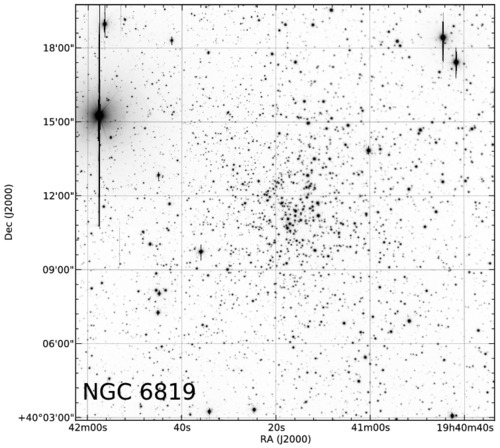}
\includegraphics[width=\hsize]{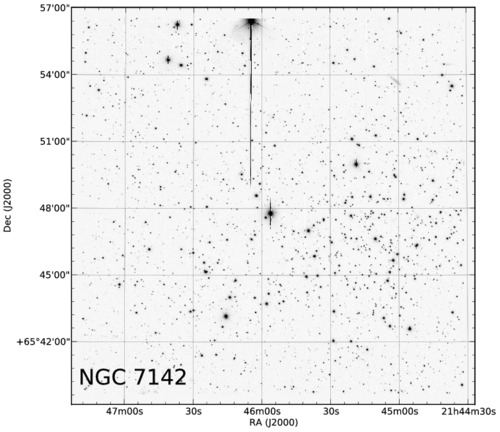}
\caption{B 300 secs mosaics of the 4 pointings for each cluster, NGC 6791, 
NGC 6819, and NGC 7142, as labelled in each panel. The field of view is 17 arcmin 
on a side. North is up, East to the left.}
\label{f1}
\end{figure}

\section{Observations and Data Reduction}
A first CCD $U,B$ observing run was carried out with the U-high-sensitive {\sc Dolores} 
optical camera mounted on the 3.6~m Telescopio Nazionale Galileo (TNG) at the Roque de Los Muchachos
Observatory of La Palma (Spain). Observations have been carried out along the three nights of 
July 29-31, 2003. {\sc Dolores} was equipped with a 2048 $\times$ 2048 pixels {\sc Loral} CCD 
with a $0^{\prime\prime}.275$ pixel size. This provided a $9^\prime.4\times9^\prime.4$ 
field of view on the sky. Four slightly overlapping fields were eventually observed across
each cluster, covering a total area of roughly $17^\prime.0\times17^\prime.0$ 
(see Fig.~\ref{f1}). The details of the observations are listed in 
Table~\ref{t1}. A further set of shallower images with 5 seconds  exposure time and
similar pointing sequence and instrumental setup has subsequently been required to recover 
saturation effects in the photometry of the brightest stars ($B \la 14$) in the fields.
These supplementary data have been kindly provided us for clusters NGC~6819 and 7142 by the TNG 
service staff along the Oct 2009 observations. Unfortunately, no useful data have been
made available for NGC~6791, so that a different correcting procedure had to be devised for this
cluster, as we discuss in Sec. 3.1.

\begin{table}
\tabcolsep 0.1truecm
\caption{Journal of observations for the 2003 run}
\begin{tabular}{lccccc}
\hline
\noalign{\smallskip}
Target& Date & Filter & Exposure & airmass & seeing \\
 &  &  & sec &  & arcsec \\
\noalign{\smallskip}
\hline
\noalign{\smallskip}
NGC~6791   & 2003 July 29   & \textit{U} & 1200   &1.02$-$1.09 & 0.8 \\
             &                                      & \textit{B} &   300    &1.01$-$1.13 & 0.7\\
PG2213+006                 &            & \textit{U} &   2$\times$30 & 1.14$-$1.62 & 0.9\\
             &                                      &\textit{B}  &   2$\times$10  & 1.14$-$1.63 & 0.8\\
NGC~6819  & 2003 July 30    & \textit{U} & 1200   &1.02$-$1.16 & 0.9\\
             &                                      & \textit{B} &   300    &1.02$-$1.18 & 0.8\\
PG2213+006                 &            & \textit{U} &   2$\times$30 & 1.14$-$1.62&0.7\\
             &                                      &\textit{B}  &   2$\times$10  & 1.14$-$1.63& 0.7\\
NGC~7142   & 2003 July 31   & \textit{U} & 1200   &1.25$-$1.28  &1.0\\
             &                                     & \textit{B}  &   300    &1.25$-$1.31 &1.0\\
PG2213+006                  &           & \textit{U} &   30     & 1.14$-$1.62  &0.9\\
             &                                      &\textit{B}  &   10     & 1.14$-$1.63 & 0.8\\
PG1525+071                  &           & \textit{U} &   30     & 1.14$-$1.62  &0.9\\
             &                                      &\textit{B}  &   10     & 1.14$-$1.63  &0.9\\
\noalign{\smallskip}
\hline
\end{tabular}
\label{t1}
\end{table}

Data have been reduced with the IRAF\footnote{IRAF is distributed by NOAO, which are 
operated by AURA under cooperative agreement with the NSF.} packages {\sc Ccdred}, {\sc Daophot}, 
{\sc Allstar} and {\sc Photcal} using the point-spread-function (PSF) method \citep{stetson87}. 
The three nights along the 2003 run turned out to be photometric and very stable, such as to allow us to 
derive calibration equations for all of the 20 observed standard stars of the 
two \citet{landolt92} fields. 

The calibration equations turned out of be in the form:
\begin{equation}
\begin{array}{rl}
 u = & U + u_1 + u_2 * X + u_3~(U-B)\\
 b = & B + b_1 + b_2 * X + b_3~(U-B),
 \end{array}
\end{equation} 
where $U,B$ are standard magnitudes, $u,b$ are the instrumental ones and $X$ is
the airmass; all the coefficient values are reported in Table~\ref{t2}.
Second order terms have also been calculated, but turned out to be
negligible ($0.005-0.015$), and therefore not included. 

In the case of NGC~6791 the specific goal of this run was to assess 
the possible presence of additional hot EHB stars {\em fainter} than $B \sim 17$,
that is the magnitude of the seven, UV-enhanced candidates originally reported by 
KU92. Quite unexpectedly, the preliminary results of these data led \citet{buson06} 
to suspect the presence of a {\it bright} EHB clump of stars
surmounting the KU92 objects. However, a closer scrutiny of the reduced data revealed that 
most of the newly detected candidates displayed in fact a too high photometric error
for their apparent luminosity and they were too close to the $B$ saturation limit 
of our deep photometry to provide 
conclusive arguments on their nature as hot sdB stars.

\subsection{Cross-check with other photometry sources in the literature}

\begin{figure}
\centerline{
\includegraphics[width=\hsize]{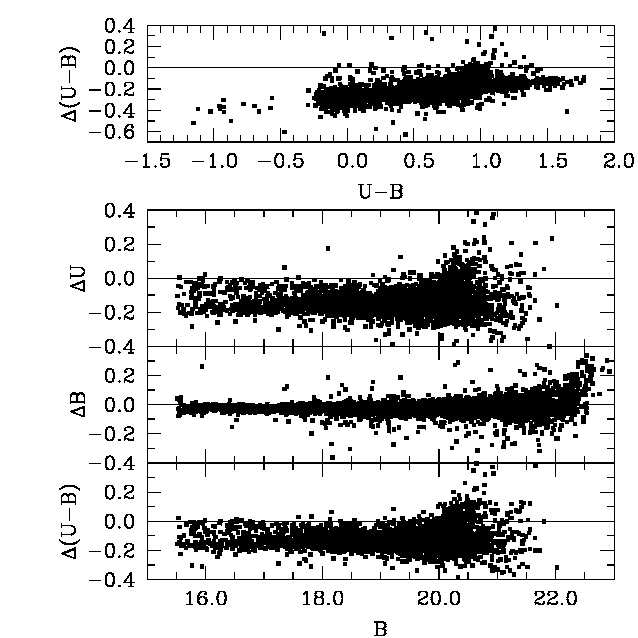}
}
\caption{$U, B$ cross-correlation of our photometry with KR95 data for cluster
NGC~6791. Color and magnitude differences for the 5510 stars in common
are displayed in the different panels versus our photometry. Mean zero-point offsets
are in the sense (``our photometry'' -- KR95). Note, in the upper panel, the 
evident $(U-B)$ color drift of KR95 photometry with respect to our data.
}
\label{f2}
\end{figure}

Cluster NGC~6791 is the only one with independent $UB$ photometry carried out by 
KR95, and this provided a valuable opportunity to check our results 
by cross-correlating the two photometric catalogs. 
The comparison restrained only to stars fainter than $B = 15.55$
mag, to safely avoid any saturation effect in our magnitude scale.
The magnitude and color residuals for the 5510 stars in common with the KR95 dataset are 
shown in Fig.~\ref{f2}. In these plots the displayed difference is in the sense 
(``our photometry'' -- KR95). As evident from the figure, a fairly good agreement is found 
for the $B$ photometry, with a mean magnitude residual $\langle \Delta B \rangle = 0.064 \pm 0.041$ 
over the whole star sample.
Major discrepancies appear, on the contrary, for the $U$ magnitudes with 
a larger zero-point offset, namely $\langle \Delta U\rangle = -0.204 \pm 0.177$, and a clear evidence 
of a color drift (see the upper panel of Fig.~\ref{f2}). One has to remind, in this regard, 
that KR95 themselves warn about possible systematics with their $U$ filter and apply an 
{\it a posteriori} offset to their $(U-B)$ color.

\begin{table}
\caption {Coefficients for standard magnitude calibration}
\begin{tabular}{lccccccc}
\hline\hline
U band & $u_1$ & $u_2$ & $u_3$ \\
\hline
Jul 29 & $0.341 \pm 0.022$ & $0.49 \pm 0.02$ & $0.103 \pm 0.033$ \\
Jul 30 & $0.349 \pm 0.018$ & $0.49 \pm 0.02$ & $0.099 \pm 0.023$  \\
Jul 31 & $0.367 \pm 0.014$ & $0.49 \pm 0.02$ & $0.139 \pm 0.018$  \\
\hline
B band & $b_1$ & $b_2$ & $b_3$ \\
\hline
Jul 29 &  $-1.544 \pm 0.010$ & $0.25 \pm 0.02$ & $\phantom{-}0.022 \pm 0.014$ \\
Jul 30 &  $-1.581 \pm 0.013$ & $0.25 \pm 0.02$ & $-0.016 \pm 0.017$ \\
Jul 31 &  $-1.581 \pm 0.012$ & $0.25 \pm 0.02$ & $-0.004 \pm 0.016$ \\
\hline
\end{tabular}
\label{t2}
\end{table}

\begin{figure}
\centerline{
\includegraphics[width=0.89\hsize]{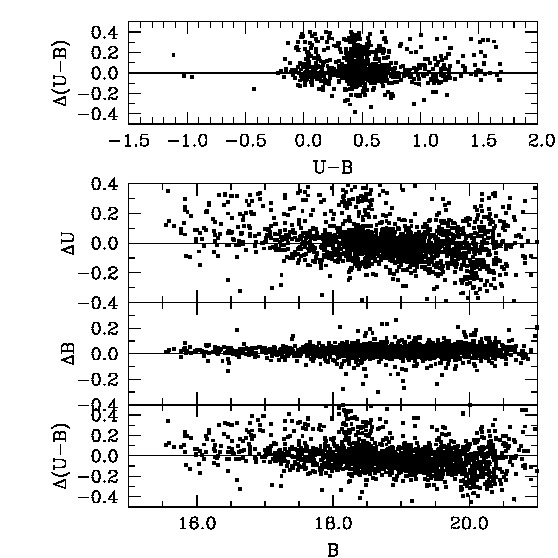}
}
\caption{Same as Fig.~\ref{f2}, but comparing with \citet{montgomery94} CCD magnitudes of 2370 
NGC~6791 stars in common with our dataset. Magnitude residuals are in the sense 
(``our photometry'' -- Montgomery et al.), and are plotted against our photometry. 
The vanishing residual distribution in the different panels confirms that our 
photometry is in the same reference as that of  Montgomery et al. }
\label{f3}
\end{figure}

An independent settlement of this apparent mismatch can be attempted by further cross-correlating
our photometry with the CCD magnitudes of \citet{montgomery94}, as shown in Fig.~\ref{f3}.
Quite comfortingly, the much smaller photometric offsets, i.e.\ $\langle \Delta B \rangle = 0.021 \pm 0.059$
and $\langle \Delta U \rangle = 0.058 \pm 0.232$, and the lack of any evident color drift for 
the 2370 stars in common confirm the excellent agreement, thus adding further strength to our 
photometry with respect to the KR95 results.

\subsection{Completeness analysis}
By looking at Table~\ref{t1}, one immediately realizes that the three clusters have been observed under the same seeing conditions. However,
Fig.~\ref{f1} shows that NGC~6791 is by far the most crowded cluster, and therefore its photometry is the most affected by crowding/incompleteness
effects. Both NGC~6819 and NGC~7142 look less affected by this problem.  
We therefore investigated incompleteness effects only on NGC~6791 images.
Completeness corrections were determined in the standard way by running artificial star experiments on the data, frame by frame, in both U and B filters. 
Basically, several simulated images were created by adding artificial stars to the original frames. The artificial stars were added at random positions and had the same color and luminosity distribution as the sample of true stars. 
To cope with potential over-crowding, up to 20\% of the original number of stars were added in each simulation. Depending on the frame, between 1500 and 2000 stars were added in this way.  The ratio of recovered to inserted stars is a measure of the photometry completeness
The results are summarized in Table~\ref{t3}, and show that both in U and  in B  the photometry has a completeness valuer larger than 50$\%$
up to 23 mag.

\begin{figure}
\includegraphics[width=0.92\hsize]{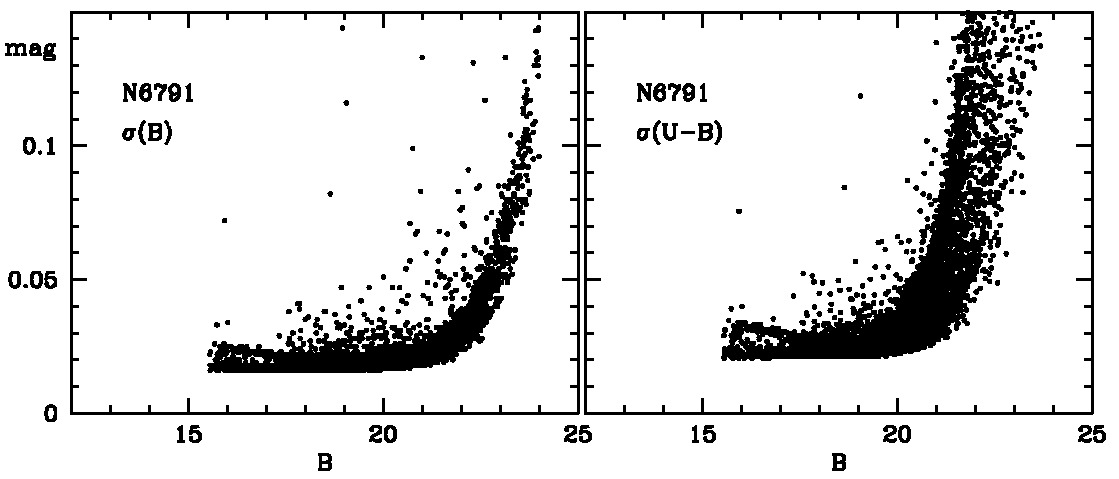}
\includegraphics[width=0.92\hsize]{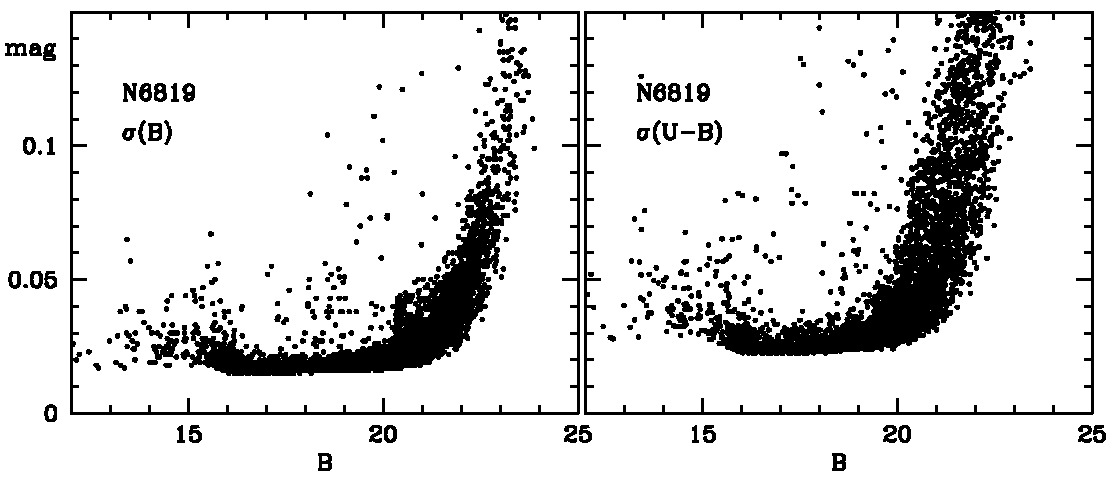}
\includegraphics[width=0.92\hsize]{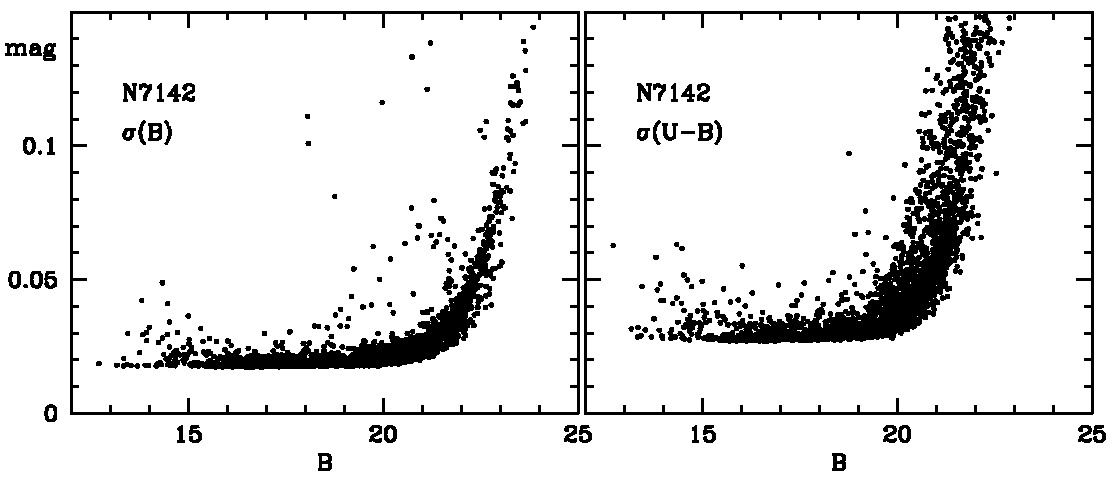}
\caption{
$B$-band internal errors from {\sc Daophot} photometry in the field of
NGC~6791, NGC~6819 and NGC~7142. Where available, ``shallow'' imagery has been used for the
photometry of the brightest ($B \la 15.5$) stars in the fields of NGC~6819
and 7142, as explained in Sec.~2. The bright-end star distribution in the NGC~6791 
field, on the contrary, has been recovered from KR95 photometry, as discussed in Sec.~3.1.
One can notice that the $B \sim 22$ mag level has been safely reached
in the three clusters, mostly within a 0.05~mag accuracy.
}
\label{f4}
\end{figure}

\begin{table}
\caption{Completeness study  for NGC 6791 as a function of the filter.}
\begin{center}
\begin{tabular}{lr r r r r }
\hline\hline
$\Delta$ Mag  &  U & B \\
\hline
13-14 &  100\% &  100\%  \\ 
14-15 &  100\% &  100\%  \\ 
15-16 &  100\% &  100\%  \\ 
16-17 &  100\% &  100\%  \\ 
17-18 &  100\% &  100\%  \\ 
18-19 &  100\% &  100\%  \\ 
19-20 &  100\% &  100\%  \\ 
20-21 &   93\% &   95\%  \\ 
21-22 &   84\% &   85\%  \\ 
22-23 &   70\% &   73\%  \\ 
23-24 &   38\% &   41\%  \\ 
\hline
\end{tabular}
\end{center}
\label{t3}
\end{table}

\section{Cluster CMDs}

\begin{figure*}
\includegraphics[width=0.507\hsize]{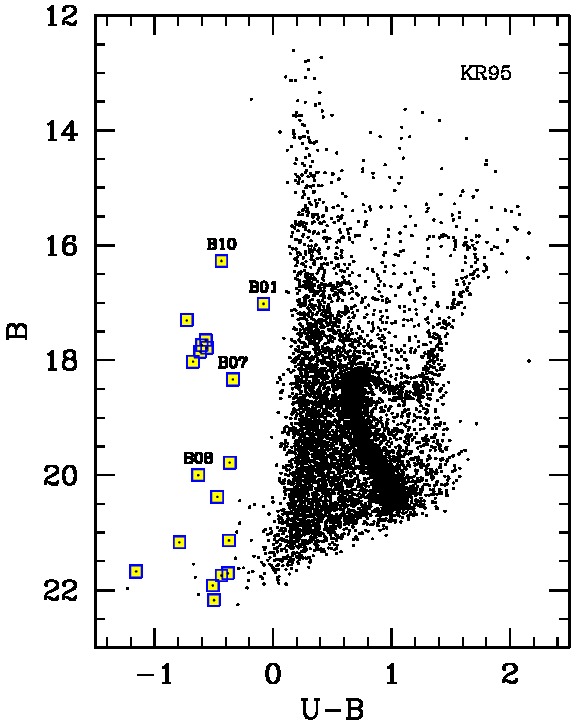}
\includegraphics[width=0.432\hsize]{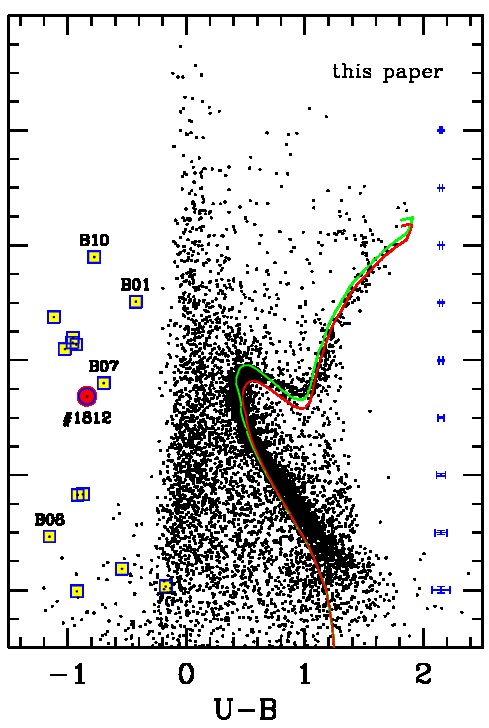}
\caption{Comparison of the $B$ versus $(U-B)$ CMD of NGC~6791 according to 
\citet{kaluzny95} {\it (left panel)} and the present paper {\it (right panel)}. 
The KR95 hot-star candidates of Table~\ref{t4} and \ref{t5} (including in particular 
the outstanding EHB stellar clump about $B \sim 18$) are marked in both plots as 
big squares. The three controversial cases of stars B01, B07 and B10 are also
labelled in the plots, together with stars B08, the hottest object in our sample. 
The big dot in the right panel indicates the new EHB candidate (ID 1812 in 
Table~\ref{t4}) we discovered in this study.
An illustrative match with the Padova isochrone set \citep{bertelli08} is displayed
in the right panel assuming for the cluster an age range between
6 and 8 Gyr and chemical mix $(Z, Y) = (0.04, 0.30)$. The theoretical models
have been shifted to an apparent $B$ distance modulus $(m-M)_B = 13.6$~mag and  
reddened by $E(U-B) = 0.13$. Typical error bars for our photometry at the 
different magnitude levels are displayed on the right.
}
\label{f5}
\end{figure*}

The {\sc Daophot} search across the field of our three clusters allowed us to confidently 
detect and measure magnitude and color for some 18,000 objects brighter than $B \simeq 24.0$ in the 
fields of the three clusters. Within these magnitude limits, the NGC~6791 sample consisted therefore
of 7774 stars, while 7683 and 3422 stars have been picked up in the NGC~6819 and 
NGC~7142 fields, respectively. A quick-look analysis of the internal photometric uncertainty of
our survey can be carried out by means of Fig.~\ref{f4}. From the plots one can appreciate that 
$B \sim 22$ mag has been safely reached throughout, mostly within a 0.05~mag accuracy.
The $B$ versus $(U-B)$ CMDs for our clusters are presented 
in the series of Figs.~\ref{f5}, \ref{f9} and \ref{f12}. 

\subsection{NGC~6791}

Our output for the NGC~6791 field is shown in Fig.~\ref{f5}, where 
we also compare with the \citet{kaluzny95} original photometry (Table~2 therein). As we were 
previously commenting on, the two datasets
exhibit zero-points differences in $U$, which make the \citet{kaluzny95} diagram 
systematically ``redder'' in $(U-B)$ color. To overcome our saturation problems
with the brightest $B$ magnitudes, however, we cross-identified all of our bright-end $B$ magnitude 
sample with the \citet{kaluzny95} catalog, and use the latter source for all $B \le 15.55$~mag
objects across our field, providing to consistently correct the \citet{kaluzny95} photometry
to our magnitude scale according to Fig.~\ref{f2}. As a result, the CMD in the right panel of
Fig.~\ref{f5} matches our own photometry for stars fainter than $B = 15.55$~mag, and the (revised) \citet{kaluzny95}
photometry, for the 110 objects brighter than this magnitude limit. 
All over, our global NGC~6791 catalog consists of 7840 entries and its resulting CMD is 
consistently the same as in the \citet{buzzoni12} analysis.
Overall, note from Fig.~\ref{f5} that our photometry turns out to be over one magnitude deeper than
\citet{kaluzny95} reaching the WD region at the faint-end tail of magnitude distribution, about $B \sim 22.5$. 

A comparison of our CMD with the YZVAR Padova isochrone set \citep{bertelli08}, as in the
right panel of Fig.~\ref{f5}, helps us constrain the overall evolutionary properties of the cluster.
For a chemical mix $(Z, Y) = (0.04, 0.30)$ the observed CMD confirms a consensus age 
between 6 and 8 Gyr \citep{anthony07,buzzoni12}, providing to shift models to an apparent $B$ distance 
modulus $(m-M)_B = 13.6$~mag, and assume a color excess $E(U-B) = 0.13$.

In the same Fig.~\ref{f5}, we encircled in both CMDs the 19 hot-star candidates 
proposed by \citet[][see Tables~1 and 2, therein]{kaluzny95}.
The sub-group of WDs is easily recognized 
fainter than $B \sim 19.5$, while an obvious EHB candidates clump stands out around $B \sim 18$.
Of these, stars B01 and B07 in the \citet{kaluzny95} original list are controversial cases 
claimed to be field stars by \citet{liebert94} according to radial velocity measurements, but 
recently re-classified as likely members of the cluster by \citet{platais11}
based on their new astrometric analysis. The case of star B10 is also a further controversial 
one as, according to \citet{kaluzny95}, this object is a blend of two stars with 
$\Delta V \sim 2$~mag and it is questioned as a likely field interloper
by \citet{platais11}. After careful inspection, object B10 can confidently be resolved in our frames,
and we are inclined to assign cluster membership at least to the brightest component of the blend.

\begin{figure}
\includegraphics[width=\hsize]{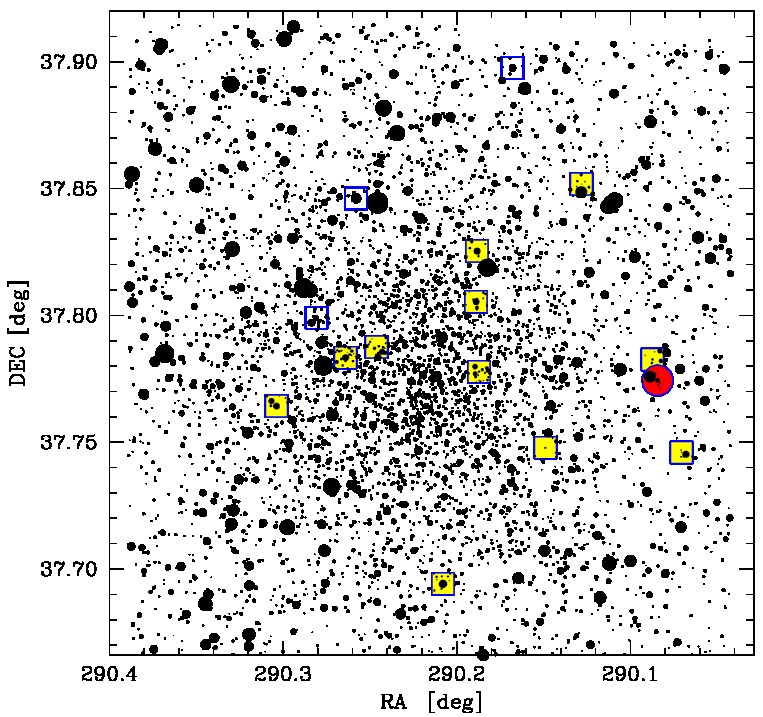}
\caption{The new EHB candidate proposed in this study is located here on the cluster map of NGC~6791
(big red solid dot) together with the \citet{kaluzny95} hot-star sample, as from Fig.~\ref{f5}
(square markers). The questioned member stars, B01 and B07 and B10, are singled out with open squares.
}
\label{f6}
\end{figure}

\begin{table}
\caption{EHB candidates in the field of NGC~6791}
\begin{tabular}{cccccc}
\hline
\multicolumn{1}{c}{ID} & 
\multicolumn{1}{c}{R.A.}        &
\multicolumn{1}{c}{DEC } &
\multicolumn{1}{c}{B} &
\multicolumn{1}{c}{(U-B)} &
\multicolumn{1}{c}{KR95}         \\
   & \multicolumn{2}{c}{(J2000.0)} &   &  & \\
\hline
\multicolumn{6}{c}{NGC 6791} \\
\hline
 377  & 19:20:40.33 & 37:53:50.9 & 16.98\tiny{(0.02)} & --0.43\tiny{(0.03)}  & B01 \\
 411  & 19:20:49.92 & 37:41:39.0 & 17.25\tiny{(0.02)} & --1.12\tiny{(0.03)} & B02 \\
 554  & 19:20:45.19 & 37:49:31.5 & 17.61\tiny{(0.02)} & --0.96\tiny{(0.04)} & B03 \\
 585  & 19:21:12.91 & 37:45:51.3 & 17.69\tiny{(0.02)} & --0.96\tiny{(0.04)} & B04 \\
 606  & 19:21:03.36 & 37:46:59.8 & 17.73\tiny{(0.02)} & --0.93\tiny{(0.04)} & B05 \\
 644  & 19:20:45.34 & 37:48:19.5 & 17.80\tiny{(0.02)} & --1.03\tiny{(0.04)} & B06 \\
1379  & 19:21:07.41 & 37:47:56.5 & 18.40\tiny{(0.02)}& --0.70\tiny{(0.04)}  & B07 \\
5939  & 19:20:35.74 & 37:44:52.3 & 21.07\tiny{(0.03)} & --1.15\tiny{(0.05)}  & B08 \\
 156  & 19:21:01.92 & 37:50:46.2 & 16.20\tiny{(0.01)} & --0.78\tiny{(0.02)}  & B10 \\
1812  & 19:20:20.22 & 37:46:27.6 & 18.63\tiny{(0.02)} & --0.84\tiny{(0.04)} & \ldots \\ 
\hline
\multicolumn{6}{l}{Notes: KR95 = ID no. from \citet{kaluzny95}.}\\
\end{tabular}
\label{t4}
\end{table}

Following \citet{buzzoni12}, one more star should be included to this EHB sample. This is target B08
in the \citet{kaluzny95} notation, the most UV-enhanced object in our sample. In spite of its 
much fainter apparent $B$ magnitude, in fact, this star is the hottest object
in our catalog, which implies a much larger intrinsic luminosity, after bolometric correction,
fully consistent with its location in the high-temperature extension of the cluster HB
(see Fig.~3 in \citealp{buzzoni12}). Star B08 partly escaped its peculiar location in
the original CMD of \citet{kaluzny95} (see left panel of Fig.~\ref{f5}) due to a redder color,
mainly in consequence of a $\sim 1$ brighter $B$ magnitude, compared to our photometry.
Such a notable difference urged a thorough check on our TNG frames to manually probe apparent
$U$ and $B$ magnitudes. A supplementary check was also carried out for star B16, which we see
$\sim 0.8$ fainter in $B$ than \citet{kaluzny95}. After careful inspection, for both cases we 
can fully confirm our magnitude estimates of Table~\ref{t4} and \ref{t5}, thus attributing 
most of the apparent discrepancy to the \citet{kaluzny95} photometry.

Overall, according to our survey, we could only detect 14 out of the 19 hot-star candidates of \citet{kaluzny95}
since 5 of them (namely B09, B12, B13, B17 and B19) happen to fall outside our field of view.
The cross-identification of the 9 \citet{kaluzny95} EHB candidates in our sample  is reported 
in Table~\ref{t4}, together with accurate J2000.0 coordinates, $B$ magnitude and $(U-B)$ 
color according to our observations. 
For reader's better convenience, the remaining 5 stars in our field are summarized in 
Table~\ref{t5}. The position of all the 14 hot stars in common with \citet{kaluzny95} is
indicated in the cluster map of Fig.~\ref{f6}.

\begin{figure}
\centerline{
\includegraphics[width=0.5\hsize]{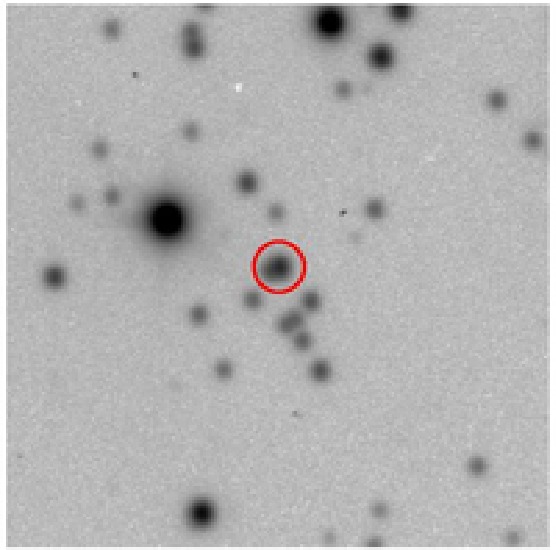}
}
\caption{The $B$-band finding charts for the new EHB candidate in NGC~6791 proposed 
in this study. This is star ID \#1812 in our catalog. Chart is $1\arcmin \times 1 \arcmin$ across, 
centered at the coordinates of Table~\ref{t4}. North is up, East to the left.
}
\label{f7}
\end{figure}

In addition to the 9 {\it bona fide} EHB stars in the \citet{kaluzny95} list, a further new 
candidate, that escaped any previous detection-- i.e. entry \#1812 in the present catalog, aka
star ``c'' in Fig.~3 of \citet{buzzoni12}-- should be added to the EHB sample.
Its dereddened $(U-B)$ color suggests for it a temperature of $T_{\rm eff} \simeq 22,300$~K 
\citep{buzzoni12}. 
This star is reported in Table~\ref{t4} and marked as a big red dot in our 
CMD of Fig.~\ref{f5} (right panel) and in the cluster map of Fig.~\ref{f6}. A more detailed 
finding chart, for future observing reference, is also reported in Fig.~\ref{f7}.
Although not confirmed spectroscopically, the projected distance 
from the cluster center makes this target compatible with its possible membership to the 
system. This statistical argument will be further detailed in Sec.~4, leading us to 
attach this star a $>70$\% membership probability. 

\begin{table}
\caption{Other cross-referenced faint hot stars in the field of NGC~6791, 
according to \citet{kaluzny95} }
\begin{tabular}{cccccc}
\hline
\multicolumn{1}{c}{ID} & 
\multicolumn{1}{c}{R.A.}        &
\multicolumn{1}{c}{DEC } &
\multicolumn{1}{c}{B} &
\multicolumn{1}{c}{(U-B)} & 
\multicolumn{1}{c}{KR95}  \\
   & \multicolumn{2}{c}{(J2000.0)} &   &  & \\
\hline
 6684  & 19:20:30.78 & 37:51:06.2 & 21.63{\tiny (0.03)}& --0.54{\tiny (0.05)} & B11 \\
 6995  & 19:20:59.08 & 37:47:15.1 & 21.94{\tiny (0.03)}& --0.17{\tiny (0.05)} & B14 \\
 4801  & 19:20:20.90 & 37:46:57.4 & 20.34{\tiny (0.03)}& --0.92{\tiny (0.05)} & B15 \\
 4784  & 19:20:44.92 & 37:46:40.2 & 20.33{\tiny (0.03)}& --0.87{\tiny (0.05)} & B16 \\
 7068  & 19:20:16.92 & 37:44:46.2 & 22.02{\tiny (0.04)}& --0.92{\tiny (0.06)} & B18 \\
\hline
\hline
\multicolumn{6}{l}{Notes: KR95 = ID no. from \citet{kaluzny95}.}\\
\end{tabular}
\label{t5}
\end{table}

Based on our revised star catalog, we also carefully reconsidered the nature 
of the striking clump of UV-strong stars, about $B \sim 15.5$, preliminarily appeared in the 
CMD of NGC~6791, as shown in \citet[][see Fig.~2 therein]{buson06}.
Although clearly detected on the deep $U$ imaging frames, these objects stand out in our 
original photometric catalog for their large $B$ photometric error, a feature that led us 
to suspect some intervening saturation effect in this band. 
For this reason an {\it ``ad hoc''} individual recognition of this bright sample on the original 
TNG images has been carried out together with an independent cross-identification 
of each target in the \citet{kaluzny95} $B$ catalog.
Our perception actually did turn out to be correct and, after recovering CCD saturation, we 
were unable to isolate in NGC~6791 any additional (clump of) UV-bright stars.\footnote{Similarly,
the saturation check also led us to reject two additional hot-star candidates of
\citet[][see labelled objects ``a'' and ``b'' of Fig.~3 therein]{buzzoni12}.} 

Overall, across our field of view, the open cluster NGC~6791 seems therefore to host a total of 
ten EHB stars.

\subsection{NGC~6819}

\begin{figure}
\includegraphics[width=\hsize]{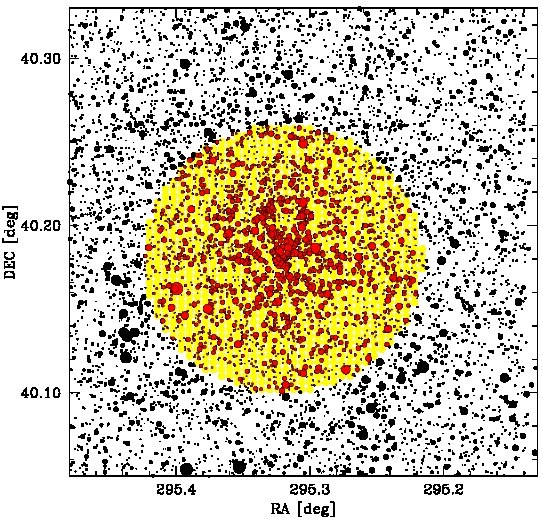}
\caption{Overall map of the surveyed field across NGC~6819. The central spot
locates the ``inner'' region of 5 arcmin radius surrounding the cluster center.
Some very bright stars East to
the cluster have been masked (see Fig.~\ref{f1}) preventing accurate photometry 
in the  relatively close region.
}
\label{f8}
\end{figure}

This study presents the first-ever $U$ CCD photometry for this cluster. Down to $B = 24.0$, our 
photometric catalog collects a total of 6504 objects. The system looks very concentrated
spatially with a substantial fraction of its 
stellar population comprised within a radius of $\sim 5\arcmin$ from the center (see
Fig.~\ref{f8}). According to the star number-density distribution, the latter can been located at
$(\alpha; \delta)_{2000.0} \simeq (19^h 41^m 17^s; +40^o~10\arcmin~47\arcsec)$. 

\begin{figure}
\centerline{
\includegraphics[width=\hsize]{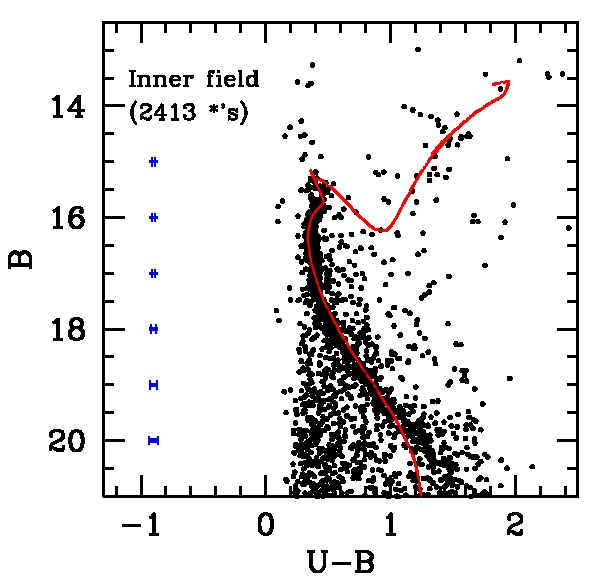}
} \centerline{
\includegraphics[width=\hsize]{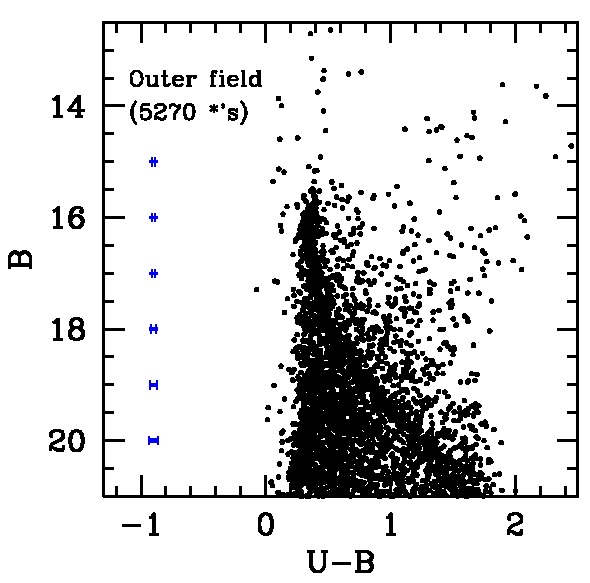}
}
\caption{{\it Upper panel:} the $B$ versus $(U-B)$ CMD of the ``inner''
region (within $5\arcmin$ from the cluster center) of NGC~6819. A total of 2413 stars
brighter than $B = 24.0$ are displayed, as labelled. Note the ``hooked'' Turn Off pattern
about $B \simeq 15.5$ and the red clump of HB stars, about 1~mag brighter, about $(U-B) \sim 1.4$.
A tentative match with the 3 Gyr Padova isochrone \citep{bertelli08} is displayed for 
$(Z, Y) = (0.04, 0.30)$. We imposed an apparent $B$ distance modulus $(m-M)_B = 12.0$~mag and  
a color excess $E(U-B) = 0.15$.
{\it Lower panel:} same plot but for stars in the ``outer'' region of the field, that is
beyond $5\arcmin$ from cluster center (5270 objects in total, within the same magnitude limit).
For both panels, typical error bars for our photometry at the 
different magnitude levels are displayed on the left.
}
\label{f9}
\end{figure}

The CMD of the 2413 stars within the 
``inner'' region (Fig.~\ref{f9}, upper panel) shows a well populated stellar main sequence (MS), 
that neatly shows up against the Galactic background. Also a red clump of HB stars,
about 1~mag brighter than the Turn Off (TO) point, is clearly visible in the figure, 
about $(U-B) \sim 1.4$. One can also notice the TO region to display an evident ``hooked'' pattern 
pertinent to stars of $M \ga 1.4~M_\odot$ growing a convective core inside. This is evocative 
of stellar populations of intermediate age.
Actually, a tentative match of the ``inner'' CMD with the Padova isochrones \citep{bertelli08}
for $(Z, Y) = (0.04, 0.30)$ (see again the upper panel of Fig.~\ref{f9}) points to an age of $\sim 3$~Gyr, 
after reddening models for a color excess $E(U-B) = 0.15$ and assuming an apparent $B$ distance modulus 
$(m-M)_B = 12.0$~mag for the cluster.

Interestingly enough, the MS stellar distribution seems to vanish toward lower luminosities 
with a clear deficiency of stars fainter than $B\sim 20$. A comparison with the observed field 
star counts, at the same magnitude level, definitely rules out any possible bias due to 
incomplete sampling and points therefore to an inherently ``flat'' (i.e. giant-dominated, in the mass range $1.02 - 1.17 M_{\odot}$) 
or truncated IMF for the cluster stellar population.

Although much more blurred and heavily perturbed by field star interlopers, all these features of
the CMD can also be recognized in the corresponding plot of the 5270 stars across the 
``outer'' region ($r > 5\arcmin$ in Fig.~\ref{f8}), as in the lower panel of Fig.~\ref{f9}. 
This clearly points to a much larger extension of the NGC~6819 system itself, as found indeed
by \citet{kalirai01}, who placed the cluster edge $\sim 9.5\arcmin$ away from the center.

\begin{figure}
\includegraphics[width=\hsize]{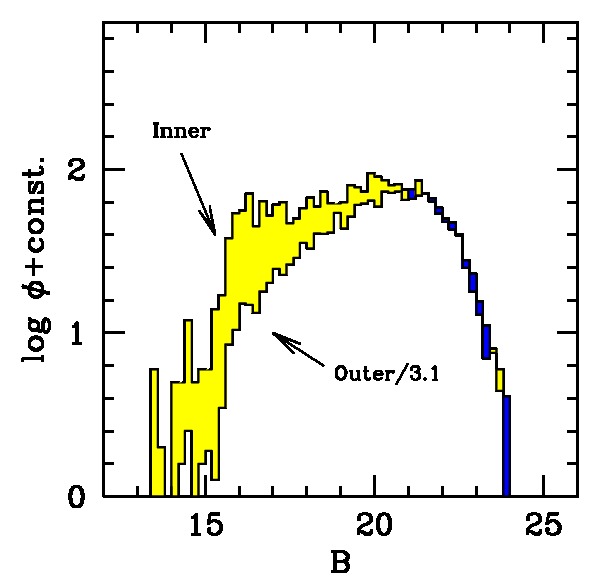}
\caption{The apparent $B$-luminosity function of the ``inner'' and ``outer'' regions
across the NGC~6819 field. To consistently compare the two regions, ``outer'' star counts 
have been reduced by a factor of $\sim 3.1$ to rescale to the same area as for the ``inner''
region.
Note the inner residual excess of bright red giants and upper-MS stars and the lack of any
central concentration for the low-MS stellar distribution fainter than $B \sim 20$.
}
\label{f10}
\end{figure}

Once rescaled to the same area across the sky, the apparent luminosity function of the
``inner'' and ``outer'' regions in NGC~6819 can consistently be compared, as in 
Fig.~\ref{f10}.
Supposing the Galaxy background to be uniformly distributed across the field, then the residual 
excess of bright red giants and upper-MS stars in the innermost
region effectively traces the cluster stellar population. In addition, the plot also confirms that
cluster low-MS stars fainter than $B \sim 20$ are spread out across the
field and do not show any central concentration. Our evidence fully supports the results of
\citet{kalirai01}, who pointed out the prevailing presence of low-mass stars ($M\la 0.65 M_\odot$)
in the outer regions of the cluster.

\subsection{NGC~7142}

As for NGC~6819, also for this cluster we are presenting here the first $U$-band CCD photometry
ever. A total of 3422 stars have been measured, brighter than $B = 24.0$.
The cluster does not clearly stand against the 
field, and looks very contaminated. This reinforces the idea that NGC~7142 is a loose open 
cluster on the verge of dissolving into Galactic disk \citep{vdb70}.

\begin{figure}
\includegraphics[width=\hsize]{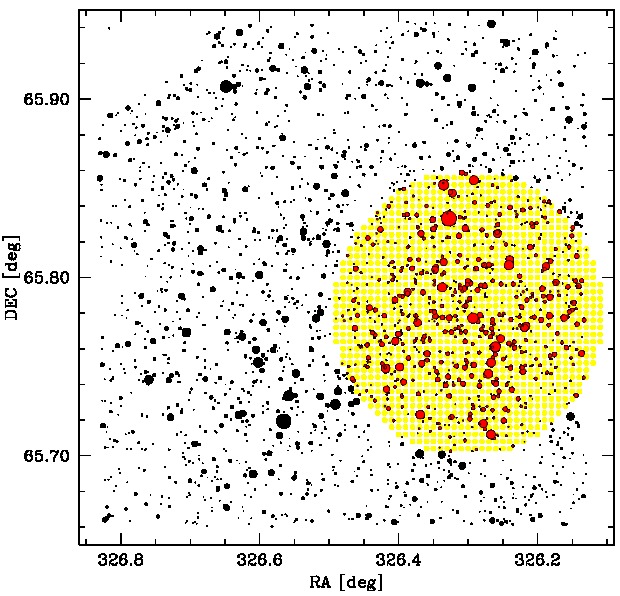}
\caption{Overall map of the surveyed field across NGC~7142. The central spot
locates the ``inner'' region of 5 arcmin radius surrounding the cluster center.
A bright star North-East to the cluster has been masked (see Fig.~\ref{f1}) thus
preventing accurate photometry in the  relatively close region.
}
\label{f11}
\end{figure}

\begin{figure}
\centerline{
\includegraphics[width=\hsize]{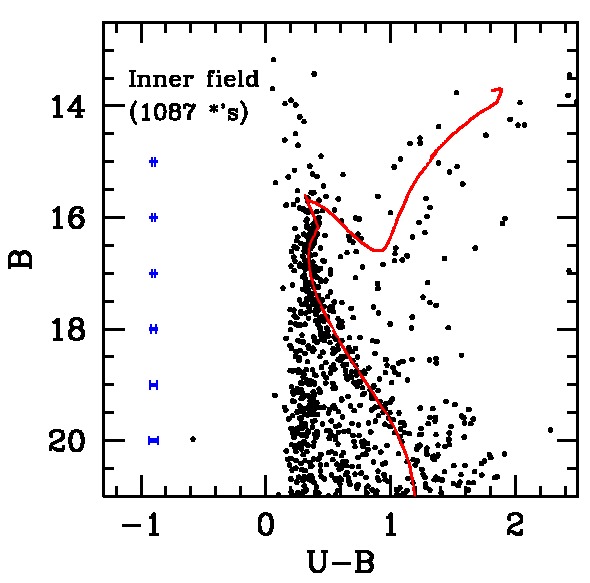}
} \centerline{
\includegraphics[width=\hsize]{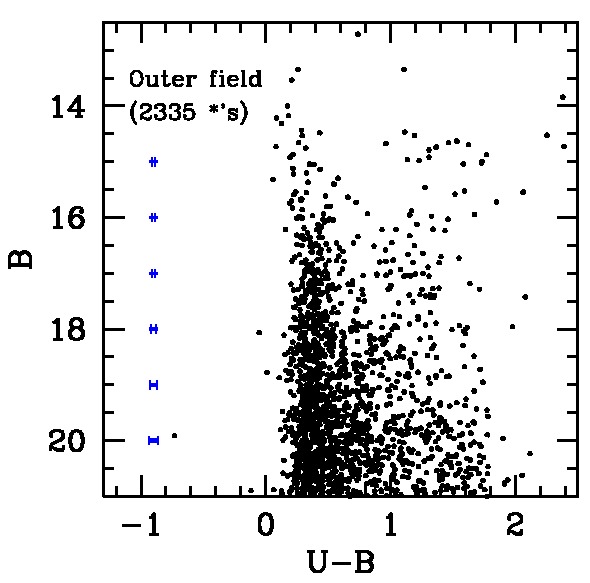}
}
\caption{Same as for Fig.~\ref{f9}, but for cluster NGC~7142. In order to enhance cluster 
visibility, the upper panel collects photometry for 1087 stars brighter than $B = 24.0$ in an 
``inner'' region within 5 arcmin from cluster fiducial center, as in the map of Fig.~\ref{f11},
while the field distribution in the ``outer'' region (2335 stars) is displayed in the lower panel.
A match is attempted in the upper panel with a 4 Gyr Padova isochrone \citep{bertelli08} for 
$(Z, Y) = (0.04, 0.30)$. An apparent $B$ distance modulus $(m-M)_B = 12.0$~mag and  
a color excess $E(U-B) = 0.10$ is assumed to rescale models.
For both panels, typical error bars for our photometry at the 
different magnitude levels are displayed on the left.
}
\label{f12}
\end{figure}

The stellar locus in the $B$ vs.\ $(U-B)$ plane can be enhanced by 
restraining our display to the densest innermost region of the system. For this reason we 
collected stars into a circular region within a $5\arcmin$ radius around the cluster center,
the latter assumed to coincide with the peak of the star number density, roughly located 
at $(\alpha; \delta)_{2000.0} \simeq (21^h 45^m 11^s; +65^o~46\arcmin~49\arcsec)$ 
(see Fig.~\ref{f11}). 
This ``inner'' sample consists of 1087 stars and evidently maximizes the fraction
of cluster members. Its CMD (upper panel of Fig.~\ref{f12}) can be contrasted with 
the ``outer'' stellar distribution across the surrounding field, amounting to a total 
of 2335 stars (lower panel of the figure).

The cluster MS neatly appears in the upper panel of the figure, with the TO point located at 
$[B, (U-B)] \sim [16.5, 0.3]$.  The MS smoothly connects with a coarse but extended RGB 
that tips about $[B, (U-B)] \sim [14.0, 2.3]$. As for NGC~6819, the diagnostic match with the 
\citet{bertelli08} Padova isochrones provides very similar results pointing however to an
age of roughly 4~Gyr (see upper panel of Fig.~\ref{f12}). As for NGC~6819, no clear 
evidence for any possible hot stellar component to be related with the cluster population seems 
to emerge from the analysis of our CMD.

\begin{figure}
\includegraphics[width=\hsize]{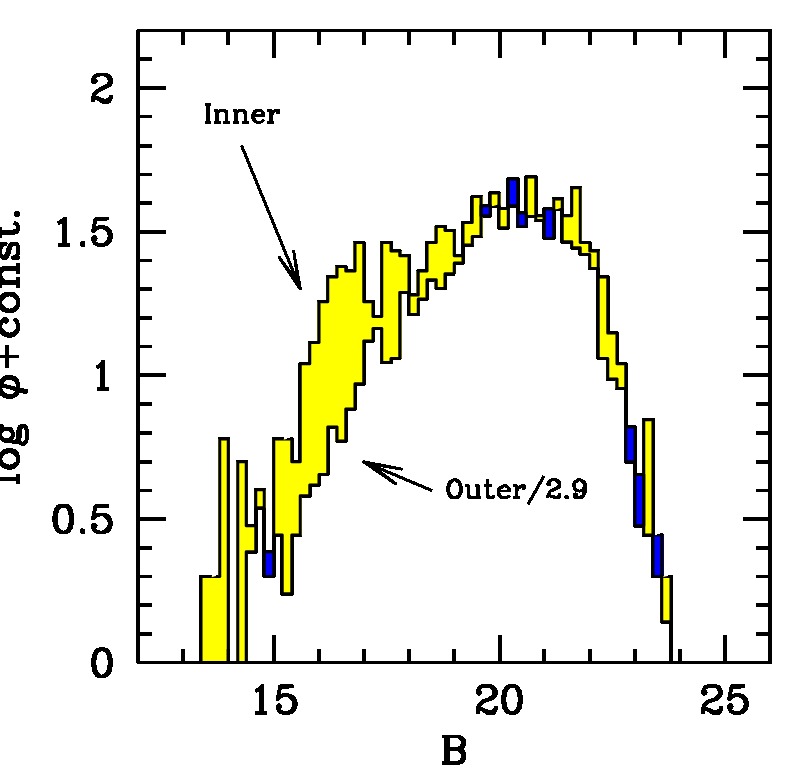}
\caption{Same as for Fig.~\ref{f10}, but for cluster NGC~7142. 
Again, to consistently compare the ``outer'' and ``inner'' areas, star counts
in the external region have been rescaled by a factor of $\sim 2.9$, as labelled in the plot.
}
\label{f13}
\end{figure}

By further extending  the comparison with NGC~6819, a notable feature one has also to remark from 
the CMDs of Fig.~\ref{f12} is an inherent deficiency of faint low-MS stars below $B \sim 20$. 
This feature becomes even more 
evident as far as the cluster luminosity function is assessed, 
although on a merely statistical basis, in terms of star count excess vs. $B$ apparent magnitude 
of the `inner' versus `outer' sample, as shown in Fig.~\ref{f13}.
Such a vanishing MS, together with the overall loose morphology of the cluster, may consistently
fit with a dynamical scenario modulated by the Galaxy interaction.
Faint low-mass stars should, in fact, be the first and most affected by Galaxy tidal stripping over 
cluster lifetime (\citet{mac08}). On the other hand, likewise NGC~6819, an apparent lack of low-mass stars could also be the tricky 
by-product of a prevailing fraction of binary (multiple?) stellar systems within the cluster population. 
If this is the case, then the entire MS locus might be affected leading, among others, to a younger 
inferred age for the cluster, as probed by a brighter TO point.

\section{Cluster membership and field star contamination}

Thanks to the large covered field, and the quite low ($b \simeq 8$-11$^o$) Galactic latitude, 
a substantial fraction of disk stellar interlopers is expected to affect our open-cluster observations.
To independently probe the Galaxy contamination along our pointing directions, and eventually assess, 
on a firmer statistical basis, cluster membership of the observed stars in each cluster, we therefore
made a Monte Carlo experiment relying on the \citet{girardi05} Galactic model to compute
synthetic CMDs of the relevant sky regions. 
To make our realizations statistically significant, we ran several trials
with varying the random seed, and then we smeared the synthetic CMDs by adding 
photometric errors as from our observations, according to Fig.~\ref{f4}. Finally, reddening at 
infinity has been applied to the theoretical $(U-B)$ colors, following \citet{schlegel98}.

\begin{figure}
\centerline{
\includegraphics[width=\hsize]{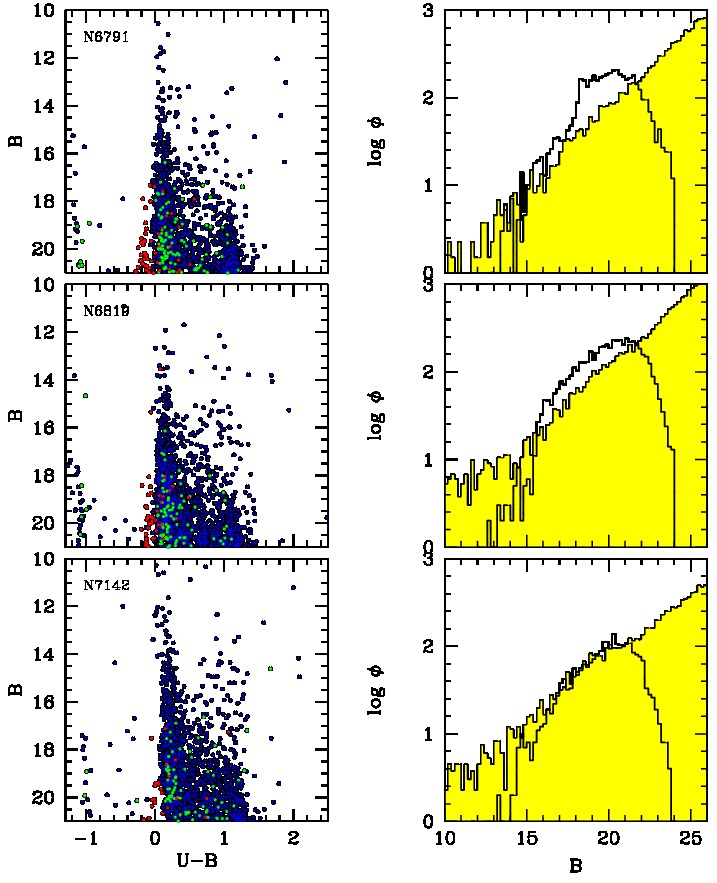}
}
\caption{Field realizations along the line of sight of NGC~6791, 6819 and 7142
simulated by means of the \citet{girardi05} Galactic model. 
The synthesis output has been scaled throughout to an area of $0.06$~square deg in order to 
consistently match the observed stellar sample of our ``outer''-field regions
(as in the maps of Fig.~\ref{f8} and \ref{f11}, for instance).
The syhntetic CMDs are displayed in the left panels, while the corresponding $B$-band luminosity 
function are computed in the right panels, and compared with our ``outer'' observations on a 
similar area of the clusters (thick-line histograms). 
Color code in the synthetic CMDs is red for halo stars, green for the thin disk, 
and blue for the thick disk. Note, all the way, the prevailing contribution of
thick-disk stars to the coarse Galactic field.
}
\label{f14}
\end{figure}

The synthetic field realizations along the line of sight of NGC~6791, 6819 and 7142
are displayed in the three left panels of Fig.~\ref{f14}, as labelled on the plots. 
To ease a direct comparison
with the corresponding CMDs of Fig.~\ref{f9} and \ref{f12}, where lower panels 
better probe the field in the off-center region of the clusters, we scaled our simulations 
to match a similar area coverage on the sky (namely $\sim 0.06$~square degrees).
The contribution of the different Galaxy components 
is color coded in the plots of Fig.~\ref{f14}, with halo stars in red, thin disk stars in green
and thick disk stars in blue. Consistently with the low Galactic latitude of our clusters, one
can notice that thick-disk stars are by far the prevailing contributors 
throughout.\footnote{According to \citet{girardi05} an exponential radial density profile is 
assumed for the thick-disk stellar component in the model, with a scale length of 2.8 kpc.}
For each cluster, in the the right panels of Fig.~\ref{f14} we also compared the $B$-luminosity functions,
as observed across the ``outer'' regions of the three fields (thick-line histograms overplotted
in each panel), with the
corresponding Monte Carlo output.\footnote{The same selection is adopted for NGC~6791, 
for which we probed the $B$ luminosity function for stars across the map of Fig.~\ref{f6} 
located $5\arcmin$ or more away from the cluster center, the latter assumed to coincide with 
the peak of the star number density at $(\alpha; \delta)_{2000.0} \simeq (19^h 20^m 52^s; +37^o~46\arcmin~13\arcsec)$.}
As expected, cluster NGC~6791 and NGC~6819 are easily recognized to ``spill over'' 
the $5 \arcmin$ region in the CMDs of Fig.~\ref{f5} and \ref{f9}
and induce a star count excess in the field luminosity function. This is not the case for
NGC~7142 which, on the contrary, seems to be fully contained within the inner $5 \arcmin$ spot
of Fig.~\ref{f11}.

Cluster membership of stars within the ``inner'' regions of our frames can be statistically assessed
by taking advantage of the Milky Way synthetic templates and relying on the \citet{mighell98} 
procedure. Restraining our test to stars brighter than $B \sim 20$ we have that, on average, about 78\% of
the objects in the ``inner'' region of NGC~6791 can confidently be cluster members. A similar figure
is obtained for NGC~6819, leading to a membership fraction of 71\% within the inner $5\arcmin$
radius. Due to its vanishing profile, the case of NGC~7142 is much worse suggesting that the
cluster actually consists of a mere 28\% of the ``inner'' stars. 

As far as the distinctive properties of the Galaxy field are concerned, at least three interesting 
differences seem to emerge from the comparison of our ``outer'' stellar samples and 
the \citet{girardi05} synthesis model of Fig.~\ref{f14}.
More specifically:

{\it i)} Even considering the smearing effect of photometric errors,
still a much broader extension toward ``redder'' colors has to be reported for our observations, with 
a larger fraction of faint ($B \ga 18$) objects exceeding $(U-B) \ga 1.5$, as shown in the
CMDs of Figs. \ref{f5}, \ref{f9}, and \ref{f12}. Distant galaxies in the background,
like high-redshift ellipticals, may be an issue in this regard as they extend 
in apparent color well redder than Galactic M-type dwarfs. However, also differential reddening effects
may give reason of this apparent discrepancy. A check in this sense has been carried out 
by relying on the relative shift of the MS locus in the cluster CMD across the field of view, as
explained in \citet{vonbraun01}. No sign of ``patched'' reddening is found across NGC~6819
and NGC~7142, although within a rough ($\Delta E(U-B) \sim \pm 0.1$~mag) internal uncertainty of 
our procedure due to poor statistics. Just a marginal (though cleaner) evidence of a reddening gradient 
appears, on the contrary, for NGC~6791, with hints for $E(U-B)$ to slightly increase 
by $\sim 0.05 \pm 0.04$~mag toward the East edge of the field.

{\it ii)} As far as the luminosity function is concerned  (see the left panels of Fig.~\ref{f14}),
the bright-end stellar distribution of the \citet{girardi05} model tends to predict all the way 
a more sizeable fraction of very bright ($B\la 13$) (thick-disk) stars, not present in the same 
amount in our stellar samples.\footnote{One has to notice, however, that we are somewhat
biased against the selection of very bright stars in our photometric catalogues.}

{\it iii)} Finally, and even more importantly, an enhanced population of WDs (of both thick- 
and thin-disk origin) is predicted in all the three fields with a clear sequence of faint
($B\ga 18$) UV-strong objects ``bluer'' than $(U-B) \sim -1$. Puzzling enough, observations
show no sign of such a sizeable field WD population, at least in the line of sight of NGC~6819 and 7142, 
while only a marginal evidence might perhaps tackle the nature of the few faintest UV stars in 
NGC~6791. 
Altogether, points {\it (ii)} and {\it (iii)} may be a hint for the \citet{girardi05} theoretical 
scheme to further tune up its assumed thick-disk morphology pointing to a shorter scale length, such as 
to reduce the overwhelming presence of  relatively close (bright) stars and their progeny of WDs in the 
solar neighborhood.

\section{Summary and discussion \label{summary}}

We reported on a multiple, UV-oriented survey in the fields of the open clusters NGC~6791, 
NGC~6819 and NGC~7142, which---owing to their super-solar 
metal content and estimated old age---represent both very near and ideal stellar aggregates to 
match the distinctive properties of the evolved stellar populations, possibly 
ruling the UV-upturn phenomenon in elliptical galaxies and bulges of spirals.
To this goal we made use of TNG $U,B$ imagery.

For each cluster, the resulting $B$ vs.\ $(U-B)$ CMD fairly well matches the fiducial  
evolutionary parameters as proposed in the recent literature, a fact that further corroborates 
the quality of our dataset. 
In particular, taking the Padova suite of isochrones as a reference \citep{bertelli08} for
TO fitting, and owing to a super-solar metallicity for all the three clusters, we confirm
for the NGC~6791 stellar population an age of $7\pm 1$~Gyr, while slightly younger figures,
i.e.\ 3 and 4~Gyr, may be more appropriate for NGC~6819 and 7142, respectively. 

\begin{figure}
\centerline{
\includegraphics[width=0.8\hsize]{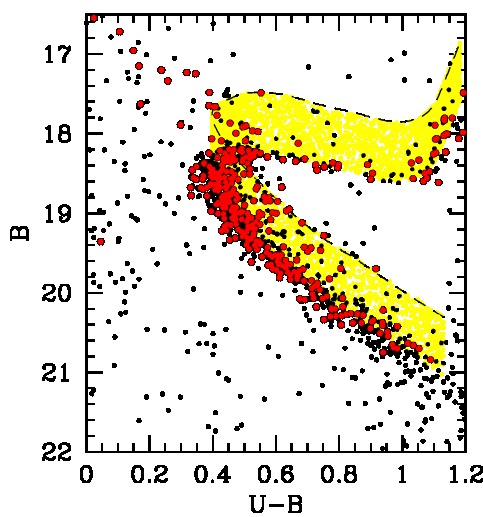}
}
\caption{A zoomed-in CMD of NGC~6791 around the MS turn-off region. 
Only stars in our catalog within a 2.5\arcmin  ~~radius from cluster center have 
been considered, in order to minimize the field-star contamination. 
According to Cudworth (2008, private communication, as cited by \citealp{twarog11})
stars with membership probability $P_m \ge 80$\% 
have been marked by big red dots. Dashed 
curve is the MS fiducial locus shifted toward 0.75~mag brighter luminosities 
such as to edge any MS+MS star pair in case of unresolved binary systems. 
See text for details.
}
\label{f15}
\end{figure}

As already pointed out by \citet{landsman98}, a bimodal HB morphology is clearly confirmed for 
NGC~6791, where the sizeable population of blue HB (BHB) stars accompanies the standard 
red clump (RHB) in a relative number partition of [BHB : RHB]~$\sim [1:4]$.
By relying on the observed HB distribution and the overall CMD morphology, a spectral synthesis of the 
cluster stellar population led \citet{buzzoni12} to emphasize the {\em unique role} of this 
NGC~6791 as a ``morceau'' of the metal-rich, evolved stellar populations characterizing the upturn-strong 
giant ellipticals. This conclusion finds out further support also by the direct experiments of 
\citet{dorman95} and \citet{buzzoni08}, where the ultraviolet spectra of the strongest 
UV-upturn galaxies, together with other integrated spectral features, like the 
H$\beta$ strength, were actually reproduced in old metal-rich stellar environments 
with a relative fraction of 20-25\% of BHB stars superposed to a canonical red HB 
evolution.
As a further piece of evidence, stemming from the analysis of the NGC~6791 CMD, one may also 
recall the recent works of \citet{bedin08} and \citet{twarog11}, where a similar figure (namely
$\sim 30\% \pm 10\%$) is independently found for the fraction of binary stars in this cluster.
Once matching the membership probability, according to Cudworth proper-motion selection
(as cited by \citealp{twarog11}) we also confirm this special feature of the NGC~6791 stellar 
population, as shown in Fig.~\ref{f15}. 
In order to minimize the field-star contamination, we restrained the stellar sample 
in the figure only to stars in our catalog within a $2.5\arcmin$ radius from cluster 
center. A ``redward-blurred'' distribution is clearly evident for the MS,
with brighter and redder outliers all nicely comprised within an upper envelope
0.75~mag brighter than the fiducial MS locus, as expected indeed for these stars 
to be MS+MS star pairs.
Such a sizeable presence of binary systems has actually been meant by \citet{bedin08}
to originate the WD peculiar distribution as observed for this cluster.
If this is the real case, then the apparent ``excess'' of EHB stars may actually
be regarded as the key connection between MS and WD evolution.

Although clearly lacking any relevant hot stellar component, clusters NGC~6819
and 7142 might add further arguments on the same line. For both systems 
a vanishing and less concentrated low-MS stellar distribution (see Fig.~\ref{f10} and \ref{f13})
could be one possible consequence of an extended presence of binary (multiple?) stellar systems 
(the lack of faintest stars being due, in this case, to their ``merging'' into brightest integrated 
objects).
Alternatively, one may call for a disruptive role of Galaxy tides on the
dynamical evolution of these open clusters, with their low-mass stars to be the most
easily stripped objects in consequence of Galaxy interaction. 

The observation of the surrounding regions along the line of sight of each cluster allows us
to usefully probe the Milky Way stellar field at low Galactic latitudes.
Our data have been tackled by the theoretical Galaxy model of \citet{girardi05},
that includes in some detail the photometric contribution of all the relevant stellar sub-structures, 
namely the spheroid system and the two thin- and thick-disk components. A match of the observed  
CMDs and $B$ luminosity functions across our fields with the theoretical predictions of the
model led us to conclude that a more centrally concentrated thick disk ( with a
scale length shorter than 2.8~kpc, as assumed by \citet{girardi05}  might 
better reconcile the lower observed fraction of bright field stars and their WD progeny.

\acknowledgments

We would like to thank the anonymous referee for a careful reading of the
draft and for a number of timely suggestions and recommendations, that
greatly helped us refining our results.
AB acknowledges the INAOE of Puebla for its warm hospitality, and the European Southern Observatory 
for awarding a visitorship to ESO premises in Santiago de Chile, where part of this work has 
been done. This project received partial financial support from the Italian Space Agency ASI, 
under grant ASI-INAF I/009/10/0 and from the Mexican SEP-CONACyT, under grant CB-2011-01-169554.

{\it Facilities:} \facility{TNG}.


\clearpage


\end{document}